\documentclass[sn-nature]{sn-jnl}

\usepackage{caption}
\usepackage{graphicx}
\usepackage{multirow}
\usepackage{amsmath,amssymb,amsfonts}
\usepackage{amsthm}
\usepackage{mathrsfs}
\usepackage[title]{appendix}
\usepackage{xcolor}
\usepackage{textcomp}
\usepackage{manyfoot}
\usepackage{booktabs}
\usepackage{algorithm}
\usepackage{algorithmicx}
\usepackage{algpseudocode}
\usepackage{listings}
\usepackage{xspace}
\usepackage[capitalise]{cleveref}
\usepackage[acronym]{glossaries}
\glsdisablehyper

\newacronym{hpo}{HPO}{hyperparameter optimization}
\newacronym{hpc}{HPC}{high performance computing}
\newacronym{hep}{HEP}{high energy physics}
\newacronym{hp}{HP}{Hyperparameter}
\newacronym{ml}{ML}{machine learning}
\newacronym{nn}{NN}{neural network}
\newacronym{pf}{PF}{particle-flow}
\newacronym{oc}{OC}{object condensation}
\newacronym{gnn}{GNN}{graph neural network}
\newacronym{llm}{LLM}{large language model}
\newacronym{mlpf}{MLPF}{machine-learned particle flow}
\newacronym{iqr}{IQR}{interquartile range}
\newacronym{lhc}{LHC}{Large Hadron Collider}
\newacronym{hllhc}{HL-LHC}{High Luminosity LHC}
\newacronym{fcc}{FCC}{Future Circular Collider}
\newacronym{cms}{CMS}{Compact Muon Solenoid}
\newacronym{lsh}{LSH}{locality sensitive hashing}
\newacronym{cld}{CLD}{CLIC-like detector}
\newacronym{hgcal}{HGCAL}{high-granularity calorimeter}
\newacronym{knn}{kNN}{k-nearest neighbors}

\newcommand{\kt}{\ensuremath{k_{\mathrm{T}}}\xspace}
\newcommand{\ptmomentum}{\ensuremath{p_{\mathrm{T}}}\xspace}
\newcommand{\ptvecmiss}{\ensuremath{{\vec p}_{\mathrm{T}}^{\kern1pt\text{miss}}}\xspace}
\newcommand{\ptmiss}{\ensuremath{\ptmomentum^\text{miss}}\xspace}

\newcommand{\TENSORFLOW} {{\textsc{TensorFlow}}\xspace}
\newcommand{\edmforhep} {{\textsc{EDM4HEP}}\xspace}
\newcommand{\fpfull}{{\textsc{float32}}\xspace}
\newcommand{\fphalf}{{\textsc{float16}}\xspace}
\newcommand{\bfhalf}{{\textsc{bfloat16}}\xspace}

\newcommand{\GEANTfour}{{\textsc{GEANT4}}\xspace}

\newcommand{\pythia} {{\textsc{Pythia8}}\xspace}
\newcommand{\ddsim} {{\textsc{ddsim}}\xspace}

\newcommand{\akfourchs}[1]{AK4-CHS\xspace}
\newcommand{\akfourpuppi}[1]{AK4-PUPPI\xspace}
\newcommand{\GeV}{\ensuremath{\,\text{Ge\hspace{-.08em}V}}\xspace}

\newcommand{\Pe}{\ensuremath{\mathrm{e}}\xspace}
\newcommand{\Pgm}{\ensuremath{\mathrm{\mu}}\xspace}
\newcommand{\PGpz}{\ensuremath{\mathrm{\pi^0}}\xspace}
\newcommand{\PGppm}{\ensuremath{\mathrm{\pi^{\pm }}}}
\newcommand{\PGpm}{\ensuremath{\mathrm{\pi^{- }}}}

\newcommand{\Pg}{\ensuremath{\mathrm{g}}\xspace}

\newcommand{\PKzL}{\ensuremath{\mathrm{K^0_L}}\xspace}
\newcommand{\ttbar}{\ensuremath{\mathrm{t\overline{t}}}\xspace}
\newcommand{\qq}{\ensuremath{\mathrm{q\overline{q}}}\xspace}
\newcommand{\zh}{\ensuremath{\mathrm{ZH}}\xspace}
\newcommand{\ww}{\ensuremath{\mathrm{WW}}\xspace}
  
\begin{document}

\title[Improved particle-flow event reconstruction with scalable neural networks for current and future particle detectors]{Improved particle-flow event reconstruction with scalable neural networks for current and future particle detectors}

\author*[1]{\fnm{Joosep} \sur{Pata}}\email{joosep.pata@cern.ch}
\author[2]{\fnm{Eric} \sur{Wulff}}\email{eric.wulff@cern.ch}
\author[3]{\fnm{Farouk} \sur{Mokhtar}}\email{fmokhtar@ucsd.edu}
\author[2]{\fnm{David} \sur{Southwick}}\email{david.southwick@cern.ch}
\author[3]{\fnm{Mengke} \sur{Zhang}}\email{mezhang@ucsd.edu}
\author[2]{\fnm{Maria} \sur{Girone}}\email{maria.girone@cern.ch}
\author[3]{\fnm{Javier} \sur{Duarte}}\email{jduarte@ucsd.edu}

\affil*[1]{National Institute of Chemical Physics and Biophysics (NICPB), R\"{a}vala pst 10, 10143 Tallinn, Estonia}
\affil[2]{European Center for Nuclear Research (CERN), CH 1211, Geneva 23, Switzerland}
\affil[3]{University of California San Diego, La Jolla, CA 92093, USA}

\abstract{
Efficient and accurate algorithms are necessary to reconstruct particles in the highly granular detectors anticipated at the High-Luminosity Large Hadron Collider and the Future Circular Collider.
We study scalable machine learning models for event reconstruction in electron-positron collisions based on a full detector simulation.
Particle-flow reconstruction can be formulated as a supervised learning task using tracks and calorimeter clusters.
We compare a graph neural network and kernel-based transformer and demonstrate that we can avoid quadratic operations while achieving realistic reconstruction.
We show that hyperparameter tuning significantly improves the performance of the models.
The best graph neural network model shows improvement in the jet transverse momentum resolution by up to 50\% compared to the rule-based algorithm.
The resulting model is portable across Nvidia, AMD and Habana hardware.
Accurate and fast machine-learning based reconstruction can significantly improve future measurements at colliders.
}

\maketitle

\section{Introduction}
One of the main approaches for event reconstruction at the \gls{lhc} is currently based on the \gls{pf} algorithm~\cite{Behrend:1982gk,Buskulic:1994wz,Abreu:1995uz,Breitweg:1997aa,Breitweg:1998gc,Bocci:2001zx,Connolly:2003gb,Abulencia:2007iy,Abazov:2008ff,CMS:2008xjf,CMS:2017yfk,Aaboud:2017aca,H1:2020zpd}, which combines measurements from different subdetectors to produce a holistic particle-based description of the entire event.
For the planned \gls{hllhc}~\cite{Apollinari:2017lan} program, due to the installation of new highly granular detector subsystems such as \gls{hgcal} for \gls{cms}, the reconstruction algorithms have to be revisited or new algorithms have to be developed to fully make use of the data which has a significantly higher complexity.
Similarly, for possible future experimental programs such as the \gls{fcc}~\cite{Selvaggi:2715344,Benedikt:2018csr}, existing algorithms have to be retuned or new ones developed, possibly many times for each new detector scenario under study.
Therefore, it is necessary to develop high-fidelity \gls{pf} reconstruction algorithms that are at the same time computationally efficient, and can be easily extended to new detector concepts without significant manual work.
Moreover, if algorithms can be found that can reconstruct events from highly granular detectors with improved fidelity, e.g. in terms of jet response, this may have important implications towards the sensitivity and thus cost-effectiveness of future experiments.

There has been considerable interest in and development of \gls{ml}-based reconstruction methods, including for \gls{pf} reconstruction.
Models for \gls{pf} reconstruction based on computer vision were investigated in~\cite{DiBello:2020bas}.
In order to train a model that reconstructs events consisting of a variable number of particles, a recipe for a loss function based on attractive and repulsive potentials was given in Ref.~\cite{Kieseler:2020wcq}.
In Ref.~\cite{Pata:2021oez}, a version of this loss was used together with a \gls{gnn}-based approach to reconstruct events with high particle multiplicity accurately.
This approach was successfully applied in the \gls{cms} experiment~\cite{Pata:2022wam,Mokhtar:2023fzl}.
Recently, Ref.~\cite{DiBello:2022iwf}, demonstrated that a network architecture based on learning a hypergraph structure can improve jet reconstruction.
In parallel, clustering using \gls{ml} has been demonstrated for high-granularity calorimeter reconstruction~\cite{CMS:2022txh}.
\begin{figure*}[htpb]
    \centering
    \caption{A conceptual overview of the machine-learned particle flow approach based on tracks, hits and clusters on one simulated \ttbar event.}
    \includegraphics[width=0.95\textwidth]{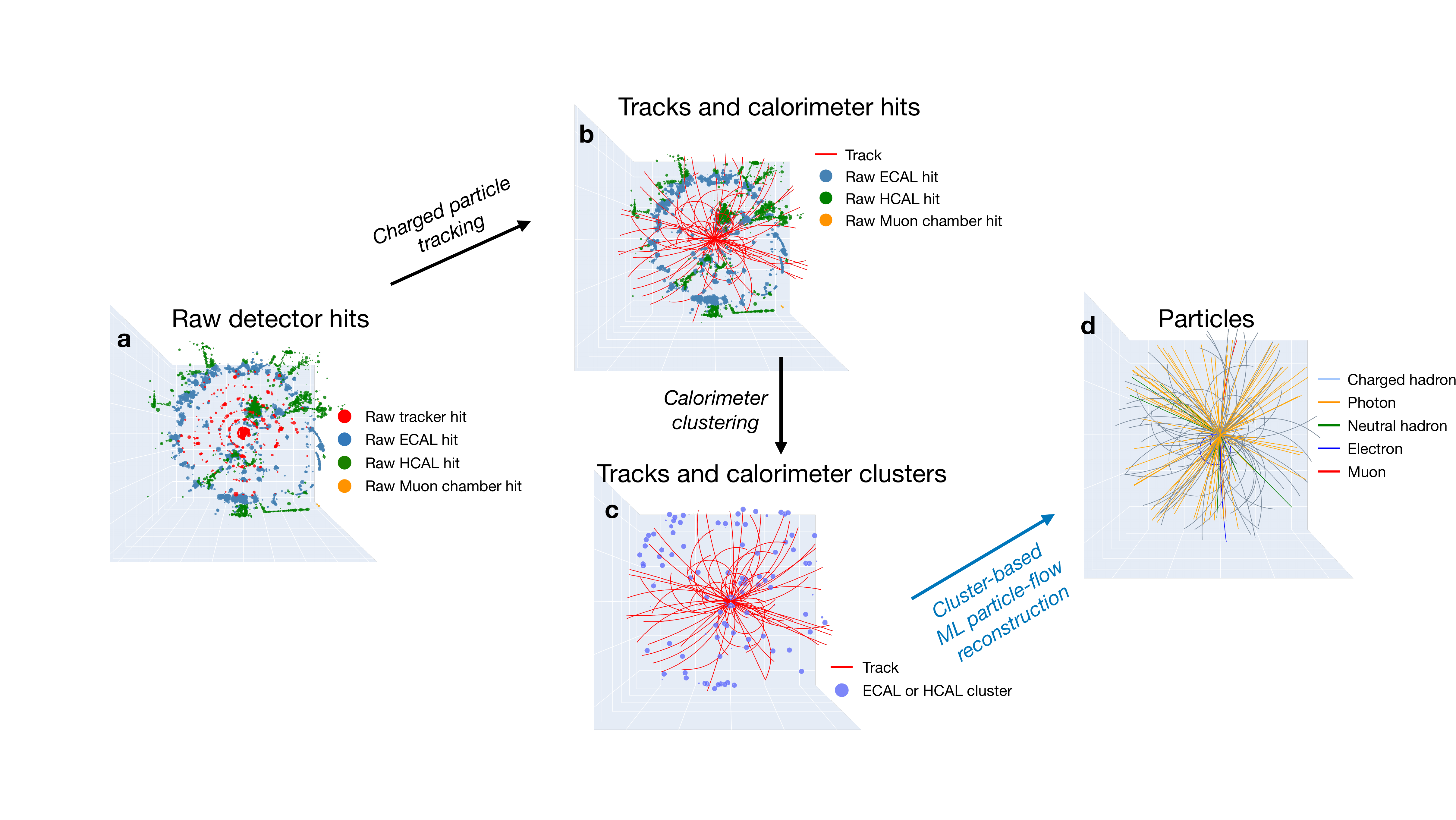}
    \label{fig:visualabstract}
    \caption*{\textbf{(a)} The raw tracker, calorimeter and muon chamber hits, embedded in position space, with the size of the marker proportional to the hit energy. \textbf{(b)} Tracking algorithms reconstruct charged particle tracks from the tracker hits, shown with their extrapolated trajectories. \textbf{(c)} The calorimeter hits are clustered to correspond better to individual particles. \textbf{(d)} The machine-learned particle flow algorithm reconstructs charged and neutral hadrons, photons, electrons and muons based on the tracks and clusters from the previous step, shown with their extrapolated trajectories.}
\end{figure*}
Extending beyond particle reconstruction, there is considerable interest and progress in reconstructing full decay trees using \gls{ml}~\cite{Kahn:2022njt,GarciaPardinas:2023pmx}.

One of the critical challenges for \gls{pf} reconstruction is the highly granular nature of the data: events can comprise hundreds of thousands of heterogeneous measurements in various detector subsystems.
This motivates studying models that can scale to large input multiplicities and efficiently process full events or batches of events simultaneously for improved throughput.
To support such studies, it is beneficial to establish open realistic simulated datasets with sufficient granularity to test various approaches.

In this paper, we utilize an open dataset of electron-positron ($\Pe^+\Pe^-$) collision events at a center of mass energy $\sqrt{s}=380\GeV$ with full \GEANTfour simulation, suitable for detector reconstruction, available in the \edmforhep~\cite{gaede2021edm4hep} format for future studies.
The specific center of mass energy was chosen due to being well studied as one of the initial proposed scenarios for CLIC and thus the baseline reconstruction software is available and tuned.
However, nothing in our proposed approach is specific to the center of mass energy or the specific detector configuration. 

We test two types of scalable \gls{ml} models as benchmarks that can process full events consisting of tens of thousands of measurements, while avoiding memory allocation or computations that scale quadratically with the input size.
Using charged particle tracks and clusters of calorimeter energy deposits, we minimize a particle-based loss function and monitor physics metrics that quantify event reconstruction performance during training.
We report the results of an extensive \gls{hpo} of the \gls{gnn}-based model performed using \gls{hpc} resources.
We then evaluate the model's physics and computational performance, as well as its portability and quantization compatibility.

We approach the challenge of high-fidelity full event reconstruction via particle flow using two alternative scalable machine learning models: kernel-based transformers and graph neural networks using locality-sensitive hashing.
After a large-scale hyperparameter optimization, we find that the \gls{gnn}-based model can reconstruct physics events with a higher degree of fidelity compared to the rule-based baseline.
At the same time, its computational cost scales linearly with the size of the input event, which is desirable for future deployment scenarios in high-granularity detectors.
Moreover, we demonstrate the portability of the model to several computational hardware processors from Nvidia, AMD, and Intel Habana.
The model can also be scaled naturally to lower-level datasets consisting of raw detector hits, if tuned tracking or clustering algorithms are not available.
Our proposed approach for particle flow reconstruction is summarized in~\cref{fig:visualabstract}.
We identify steps for future development and also publish the code and datasets following the findable, accessible, interoperable, and reusable (FAIR) principles~\cite{fair,Chen:2021euv,Duarte:2022job} for reproducibility and future development.

\section{Methods}
In this section, we describe the loss function for particle flow reconstruction, the specific \gls{nn} approaches as well as the procedure for dataset generation.

\subsection{Loss function}
The optimization goal for full event particle reconstruction follows the \gls{mlpf} approach previously used in Refs.~\cite{Pata:2021oez,Pata:2022wam,Mokhtar:2023fzl}, and applies a physics-inspired ansatz to simplify the event-based reconstruction loss to a particle-based classification and regression loss.

The input to the model is a set of detector elements: charged tracks and calorimeter clusters (or alternatively, raw calorimeter hits), each described by a feature vector $\mathbf{x} \in X$.
The input features for the tracks consist of the track \ptmomentum, the pseudorapidity of the particle momentum (track tangent) $\eta$, the azimuthal angle of the particle momentum $\phi$, the track goodness-of-fit $\chi^2$ and the number of degrees of freedom $N_\mathrm{dof}$, the slope of the track $\tan{\lambda} = p_z / \sqrt{p_x^2 + p_y^2}$, the signed impact parameter $D_0$ with respect to the origin $(0,0,0)$ at interaction point in the $xy$ plane, the signed curvature of the track $\Omega = \mathrm{sign}(q)/R$ defined via the track radius $R$ and charge $q$, the position of the track along the $z$-axis $Z_0$ with respect to the origin~\cite{Kramer:2006zz}.
For calorimeter clusters, the features consist of transverse energy $E_\mathrm{T}$, $\eta$, $\phi$, electromagnetic calorimeter energy $E_{\mathrm{ECAL}}$, hadronic calorimeter energy $E_{\mathrm{HCAL}}$, Cartesian spatial coordinates $x$, $y$, $z$; the number of hits and the cluster size, measured in standard deviations of the hit sets in $x$, $y$, and $z$.

The target of the model is a set of stable particles such as charged and neutral pions, photons, electrons, and muons, defined through sufficient energy matched to detector hits and with a \pythia8 status code 1.
This specific choice of target particles represents a reasonable optimization target, but may not be optimal for all cases.
The definition of an optimal ground truth for particle reconstruction is left to future work.
Each particle is described by a particle type, charge, and continuous four-momentum values, combined to form a target feature vector $\mathbf{y} \in Y$.
The goal of the model is, for each event, to reconstruct the set $Y$, given the set $X$, i.e. $\Phi(X) \rightarrow Y' \simeq Y$.
To make the set-to-set translation tractable with \gls{ml}, and to be able to treat particle multiplicities in the range of $|X|, |Y| \simeq 10^4$ with a single model pass, each target particle is associated with a single unique input element.

The target particles are assigned to input elements by an injective, non-surjective function based on a physics prior.
The general approach is based on the \gls{oc}~ \cite{Kieseler:2020wcq}, which includes a potential between inputs and outputs.
Here, we only associate charged particles to tracks and neutral particles to the highest-energy calorimeter cluster (or hit), as a tradeoff between expressiveness and computational cost.
The event reconstruction loss can be written as a sum over the input elements in each event

\begin{equation}
L(Y,Y') = \sum_{i} L_\mathrm{cls}(y_i, y'_i) + L_\mathrm{reg}(y_i, y'_i).
\end{equation}

Here, $L_\mathrm{cls}(y_i, y'_i)$ is a classification loss between the predicted and target particle type and charge labels. 
The classification task is imbalanced, so we use the focal loss~\cite{lin2017focal} for particle type classification, which assigns greater weight to samples that are difficult to classify.
Similarly, $L_\mathrm{reg}(y_i, y'_i)$ is a regression loss for the momentum components, where we use the Huber loss~\cite{huber} to reduce the effect of outliers.

We note that in the recent hypergraph-based reconstruction method~\cite{DiBello:2022iwf}, the strict particle-to-element association is completely avoided, and the associations are instead made an optimization target for an intermediate model.
This hypergraph reconstruction approach shows excellent physics performance on small jet-based samples, but further work is needed to extend it to full events with large particle multiplicities and to demonstrate computational feasibility on realistic datasets.

\subsubsection{Scalable neural network models}
We now move to the specific \gls{nn} structure of the reconstruction model $\Phi(X)$.
It could be implemented with a simple feedforward network over the individual feature vectors of each input element, i.e. $\phi(x) \rightarrow y'$.
However, such a model would be unable to consider correlations between related input elements, such as tracks and calorimeter clusters (or hits) arising from a single particle.

An early approach proposed for full-event reconstruction previously used a \gls{gnn} with a learnable graph structure between the input nodes~\cite{Pata:2021oez}.
This is conceptually similar to a transformer with full self-attention, which has also recently been thoroughly investigated~\cite{DiBello:2022iwf}.

However, the evaluation of full self-attention can be computationally demanding and prohibitive for large input multiplicities (${\geq}10^{4}$)~\cite{wang2020linformer}.
Such input multiplicities can be easily reached when considering all tracks and calorimeter clusters in future high luminosity collider events.
It is possible that the local nature of the problem can be exploited directly by first pre-partitioning the event and running the full attention-based model only on subsets.
However, this requires careful implementation to avoid artifacts from partitioning and subsequent stitching.

Here, we study models that can naturally scale to large input multiplicities by avoiding pairwise memory allocations or computation.
Increasing the input multiplicity that \gls{ml} models can process simultaneously without manual splitting also receives considerable attention in the literature because of its wide range of applications~\cite{NEURIPS2020_1457c0d6,touvron2023llama}, and our approaches are based on existing research:
\begin{itemize}
\item The first model uses a dynamically learned graph structure~\cite{Pata:2021oez}, but avoids a full quadratic allocation or computation by using a learnable binning based on \gls{lsh} in each graph building layer~\cite{Pata:2022wam, Mokhtar:2023fzl}, inspired by the \texttt{Reformer} architecture~\cite{kitaev2020reformer}.
This approach divides each event into bins of fixed size based on a learnable function.
\item The alternative model is a kernel-based transformer in which the softmax self-attention layer is approximated using positive orthogonal random features~\cite{choromanski2020rethinking}.
This approach uses a mathematical approximation based on random projections to avoid computing the full attention.
\end{itemize}

Both of these approaches avoid memory allocations or computations that are quadratic in input multiplicity.
The \gls{lsh}-based approach was first used in \cite{Pata:2021oez}, initially being paired with a \gls{knn} graph structure in each bin.
In this paper, we instead use an all-to-all fully connected graph in each bin, as we have found it to significantly outperform the \gls{knn} based approach computationally as well as in terms of fidelity.
Moreover, in this paper, we compare the \gls{lsh}+\gls{gnn} approach with the kernel-based transformer alternative systematically.
Thus, this paper offers the first detailed comparison of alternative scalable models for particle flow reconstruction.

\begin{figure*}[htpb]
    \centering
    \caption{One layer of the locality sensitive hashing based graph neural network.}
    \includegraphics[width=0.95\textwidth]{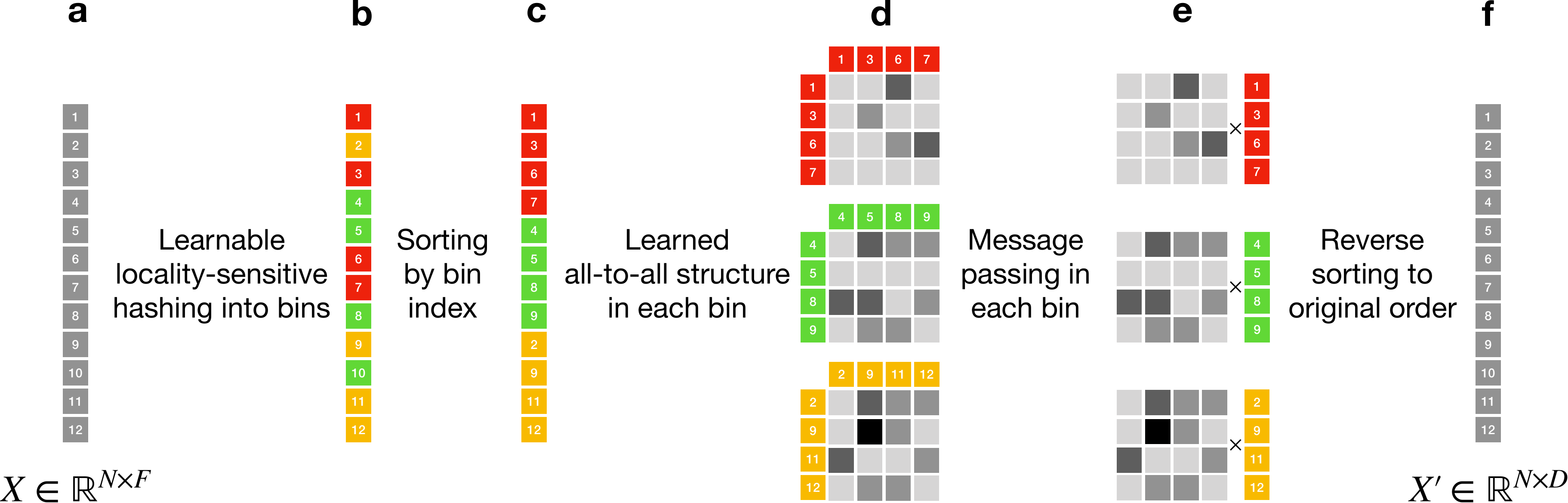}\\
    \caption*{
    \textbf{(a)} The input event of $N$ tracks and clusters (elements) is described by a list of $N$ feature vectors, one $F$-dimensional vector per element.
    \textbf{(b)} Each element is first assigned into a bin based on a learnable function, denoted with the color of the box.
    \textbf{(c)} The elements are then sorted according to the bin, such that elements in the same bin are consecutive.
    \textbf{(d)} In each bin, a full all-to-all learnable adjacency matrix is constructed between all the elements in the bin, with the learned element-to-element association illustrated by the color of the matrix element.
    \textbf{(e)} This matrix is used for message passing in each bin, multiplying the corresponding bin feature vectors with the learned adjacency matrix.
    \textbf{(f)} The output is a list of $N$ transformed feature vectors, one $D$-dimensional vector for each track or cluster.
    Each input and output vector is represented by a single gray box, the order of the input and output feature vectors is preserved.
    }
    \label{fig:visualabstract_models}
\end{figure*}

The \gls{lsh}-based \gls{gnn} is constructed as follows.
A full all-to-all graph between $N$ input elements (tracks and clusters) in the event would have dimensionality $N^2$, which for an event with $N=10^3$ tracks and clusters would require $N^2 = 10^6$ individual associations to be stored and computed for each layer.
Instead, we split the event dynamically into bins with a fixed and constant size $B \ll N$, and define the element-to-element connectivity only in each bin, each bin requiring $B^2$ associations.
For an event with $N$ input elements, the number of bins is then determined dynamically at runtime by $N_B = N/B$, with the last bin being padded if necessary.
Therefore, instead of computing an $N \times N$ adjacency matrix for each layer, we instead compute a three-dimensional $N_B \times B \times B$ adjacency matrix.
This means that an event with $N=10^3$ elements and bin size $B=10^2$ would consist of $N_B = 10$ bins, requiring $B N_B^2 = 10^5$ instead of $N^2 = 10^6$ individual associations.
The graph structure defined by the association matrix consists of $N_B$ disjoint graphs of $B$ elements each, which is not a major limitation because the graph building layers can be stacked multiple times, each building different disjoint graphs.
From a physics point of view, it is reasonable to expect that nearby (for some learnable definition of neighborhood) inputs should have a stronger association than input elements in highly separated parts of the detector.
This graph building layer can be implemented using elementary matrix operations in a fully batched and differentiable way using \TENSORFLOW~\cite{tensorflow2015-whitepaper,tensorflow_developers_2023_7753622}.
This is similar to manually partitioning and restitching the event, but achieved directly in the \gls{nn} model using fully differentiable operations, rather than with a manual heuristic.
The \gls{gnn} is based on stacked layers of graph building and convolution, with the number of layers being a configurable hyperparameter.

The alternative kernel-based transformer model avoids quadratic scaling using the following approach.
For $N$ elements, given queries $Q\in \mathbb{R}^{N{\times}d_q}$ and keys $K\in \mathbb{R}^{N{\times}d_k}$, the attention mechanism encodes a value matrix $V\in \mathbb{R}^{N{\times}d_v}$ as
\begin{equation}
\mathrm{Attn}(Q,K,V) = \mathrm{softmax}\left(\frac{QK^\intercal}{\sqrt{d_k}}\right)V.
\end{equation}
Here, the $\mathrm{softmax}(QK^\intercal)$ operation creates a full $N{\times}N$ matrix.
As in Ref.~\cite{choromanski2020rethinking}, we define a transformation $\psi(\mathbf{x}) \rightarrow \mathbf{x'}$ that transforms an input feature map $\mathbf{x}$ using predetermined random projections to a new feature space $\mathbf{x'}$.
For a sufficiently large number of random projections, attention can be approximated as
\begin{equation}
\mathrm{Attn}(Q,K,V) \simeq Q' (K'^\intercal V)
\end{equation}
where $Q'$ and $K'$ are the query and key matrices after the random feature mapping $\psi$, respectively.
Allocation of the entire $N{\times}N$ matrix is avoided, since the order of operations is changed to first multiply keys with values and then subsequently with queries.
In the special case of self-attention, $Q$, $K$, and $V$ are all derived from $X$ through a linear layer, and the self-attention mechanism can be seen as an analogy to graph building and message passing.
A visual overview of the supervised learning setup, as well as the \gls{lsh} model structure, is shown in \cref{fig:visualabstract_models}.

The complete model for \gls{pf} reconstruction is then implemented based on stacked layers of \gls{lsh}-based \gls{gnn} or self-attention and feedforward networks, with the number of layers being a \gls{hp}.

\subsubsection{Dataset}
Having defined \gls{nn} models for \gls{pf} reconstruction that avoid $\mathcal{O}(N^2)$ memory allocation and computational scaling, we train and test them on a realistic dataset.
We generate $\Pe^+\Pe^-$ collision events with \pythia (v8.306)~\cite{Bierlich:2022pfr} and carry out a complete detector simulation with \GEANTfour (v11.0.2) using the \texttt{Key4HEP} framework (v2023-01-15) \cite{ganis2022key4hep}.
In particular, we use the Compact Linear Collider (CLIC) detector model~\cite{CLICdp:2017vju,CLICdp:2018vnx}, along with the \texttt{Marlin} reconstruction code~\cite{Gaede:2006pj}, and the \texttt{Pandora} package~\cite{Marshall:2012hh,Marshall:2012ry,Marshall:2015rfa} for a baseline \gls{pf} implementation.
The CLIC detector model is chosen because it is publicly available, well documented and realistic, and similar to detector concepts that are in use at \gls{lhc}, or under consideration for either \gls{hllhc} or \gls{fcc}.

The CLIC detector model is based on the CMS detector at CERN.
It features a superconducting solenoid with an internal diameter of 7\,m, providing a magnetic field of 4\,T in the center of the detector.
Silicon pixel and strip trackers, the electromagnetic (ECAL) and hadron calorimeters (HCAL) are embedded within the solenoid.
Each subdetector is divided into a barrel and two endcap sections.
The ECAL is a highly granular array of 40 layers of silicon sensors and tungsten plates. 
The HCAL is built from 60 layers of plastic scintillator tiles, read out by silicon photomultipliers, and steel absorber plates. 
The muon system surrounding the solenoid consists of 6 and 7 layers of resistive plate chambers interleaved with yoke steel plates in the endcap and barrel respectively.
Two smaller electromagnetic calorimeters, LumiCal and BeamCal, cover the very forward region of the detector on either side of the interaction point~\cite{CLICdp:2017vju,CLICdp:2018vnx}.

Collision events are generated with different physics processes to test the out-of-distribution performance of the model.
In particular, we use \pythia to generate ${\simeq}10^6$ inclusive \ttbar, \zh, fully hadronic \ww each, and ${\simeq}2\times10^6$ \qq events.
Furthermore, we generate ${\simeq}7\times10^5$ of \ttbar PU10 events with a beam-induced hadronic overlay $\Pg\Pg \to \qq$ interactions, corresponding to an average of 10 additional interactions per event, known as pileup (PU), to test the stability under varying conditions.
We also generate single $\Pe^\pm$, $\Pgm^\pm$, $\PKzL$, $\PGpz$, $\PGppm$, neutron, and photon particle gun samples, with a uniform energy distribution in $E\in[1, 100]\GeV$, generated using the \ddsim~\cite{petrivc2017detector} package, $\simeq 10^6$ events each for performance tests.
The \qq and \ttbar were split in an 80/20 ratio to form train/test samples, while the \zh, \ww, \ttbar PU10 and single particle gun datasets were never used in training.

The datasets with generator particles; reconstructed tracks, hits and calorimeter clusters; as well as reconstructed particles from the baseline \texttt{Pandora} algorithm are saved in the \texttt{EDM4HEP} format, including all the relevant associations.
Overall, the size of the dataset is approximately 2.5\,TB before preprocessing to the \gls{ml}-specific format using the \textsc{tfds} library~\cite{TFDS}.
The raw datasets in \texttt{EDM4HEP} format, along with the scripts and configurations to generate the data, are available at \cite{pata_joosep_2023_8260741}.

There are about 50--500 tracks or calorimeter clusters per event, while the multiplicity for raw calorimeter hits is considerably larger at 5--$15 \times 10^3$ per event.
In this paper, we analyze the datasets at the level of the tracks and calorimeter clusters to study the physics performance of models.
We note that it is straightforward to apply the model on tracks and raw calorimeter hits, however, we leave this analysis for a follow-up paper.
The input elements and target particles have different underlying signatures in different samples, depending on the physics process, therefore the models will have to demonstrate out-of-distribution generalization.

Jets are defined by clustering particles with the generalized \kt algorithm ($R=0.7, p=-1$) for $\Pe^{+}\Pe^{-}$ colliders~\cite{Cacciari:2008gp, Cacciari:2011ma} with a minimum $\ptmomentum \ge 15$\GeV.
No additional quality cuts are applied on the jets at this stage.
We also evaluate the \ptvecmiss and total 3D momentum of the reconstructed and generated particles.
The missing transverse momentum vector \ptvecmiss is calculated as the negative vector sum of the transverse momenta of all particles in an event, and its magnitude is denoted as \ptmiss.
We use these quantities to assess the performance of \gls{mlpf} and the baseline \gls{pf} reconstruction with respect to the ground truth defined by the generator particles.

\section{Results and discussion}

\begin{figure*}[htpb]
    \centering
    \caption{The model validation loss and physics performance throughout training for the graph neural network and kernel-based transformer, before and after hypertuning.}
    \includegraphics[width=0.95\textwidth]{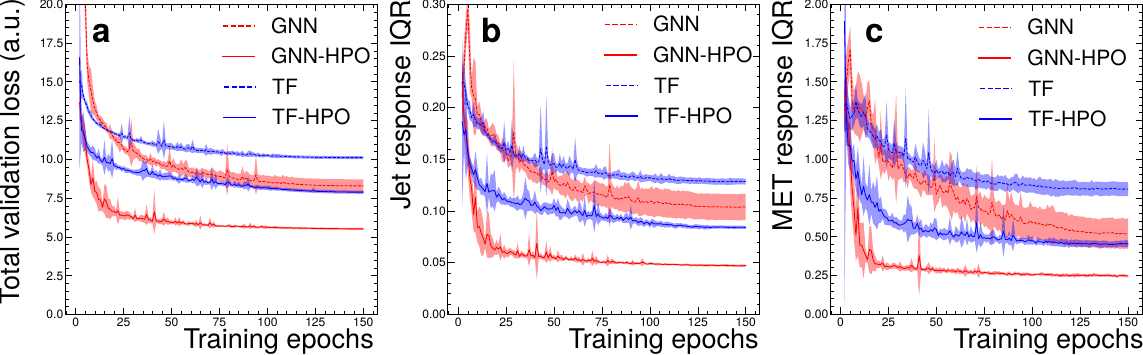}
    \caption*{The validation loss \textbf{(a)}, jet  \textbf{(b)} and \ptmiss resolution \textbf{(c)}, parameterized by the interquartile range (IQR) of the response distribution, for the graph neural network (GNN, red) and kernel-based transformer (TF, blue) models before (dashed lines) and after hyperparameter optimization (HPO, solid lines), evaluated using 10 trainings, each trained on four Nvidia A100 devices in a distributed data-parallel manner.
    Lower values correspond to better physics performance.
    Only the loss is explicitly minimized, the jet and \ptmiss resolution improvement emerges from the minimization of the loss function.
    We show the evolution of these quantities during the full training process, where only the random seed differs in each run.
    The shaded regions show the standard deviation of the metrics across the runs.}
    \label{fig:hypertuning}
\end{figure*}

Several steps were taken to ensure efficient training on large-scale datasets.
The models can be implemented using matrix operations in native \TENSORFLOW.
Given the varying size of events, the model implements variable-sized batching, supporting batching events at fixed batch size increments, with the batch size inversely proportional to the size of the events in the batch.
This was necessary to reduce the amount of zero-padding, and to enable automatic kernel compilation, so that a limited number of kernels have to be compiled for each bucket size increment. 
Since the total size of the datasets can reach several terabytes, efficient larger-than-RAM training is enabled through dynamic dataset loading and interleaving using the \TENSORFLOW \textsc{datasets} library.
The model supports mixed-precision training in \bfhalf and \fphalf.
The training in \bfhalf results in loss values that are stable and largely compatible with that of \fpfull, while we find that the dynamic loss scaling for \fphalf results in \textsc{NaN} gradients and does not converge.
At the time of writing, software support for \bfhalf in \TENSORFLOW is limited, such that the operations are not placed on tensor cores and the use of \bfhalf does not result in increased training throughput, while training with \fphalf results in moderate speedups of 30--40\% for the \gls{gnn} model.
Improving training throughput using \bfhalf is the focus of future work.

An extensive \gls{hpo} was performed for both the \gls{gnn}- and kernel-based transformer model using the JURECA supercomputer~\cite{jureca}.
JURECA is a pre-exascale modular supercomputer operated by J\"{u}lich Supercomputing Centre at Forschungszentrum J\"{u}lich.
The system consists of a flexible Data Centric (DC) module, based on the Atos BullSequana XH2000.
It has, among others, 192 accelerated compute nodes with four NVIDIA A100 GPUs and two AMD EPYC 7742 CPUs each.
We employ a similar approach to \gls{hpo} as in Ref.~\cite{Wulff_2023}, using Bayesian Optimization in combination with the ASHA~\cite{asha} algorithm.
The optimization was performed in a distributed manner on 96 GPUs spread over 24 compute nodes, consuming roughly 7000 and 5000 GPU hours for the optimization of the kernel-based transformer and the \gls{gnn} model respectively.
The hyperparameter optimization details can be found in Supplementary Note 1: Hyperparameter optimization.

The performance improvements achieved from \gls{hpo} are presented in \cref{fig:hypertuning}.
We find that the optimized versions of both the \gls{gnn} and the kernel-based transformer significantly outperform the unoptimized versions.
The relevant evaluation criterion for model selection is the reconstruction of jets and \ptmiss compatible with those at the generator level, a target that is not explicitly trained for, but can be reached by minimizing the particle-level loss.
The optimized version of the \gls{gnn} significantly outperforms the kernel-based transformer, although both use a similar number of trainable parameters (${\simeq}5\times10^6$).
As the \gls{gnn}-based model has significantly improved validation loss and physics performance, we focus on it for the rest of this paper.

\begin{figure*}[!h]
  \centering
   \caption{Performance of the particle flow (PF) and machine-learned particle flow (MLPF) algorithms on single particle gun samples.}
  \includegraphics[width=0.95\textwidth]{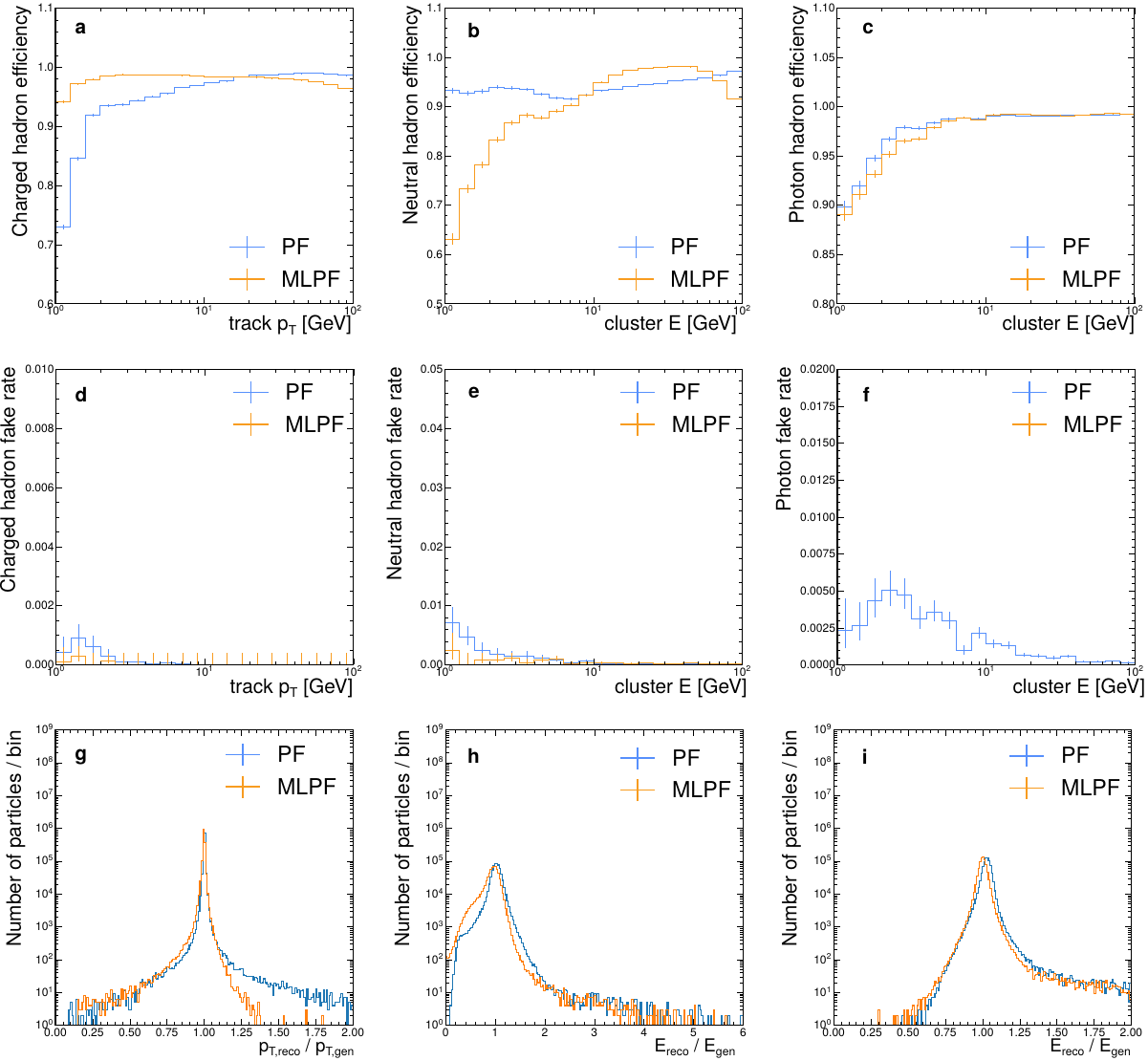}
  \caption*{Charged hadron efficiency in \textbf{(a)}, neutral hadron efficiency in \textbf{(b)}, photon efficiency in \textbf{(c)}. The efficiencies are parameterized as a function of track \ptmomentum or cluster energy.
  The charged hadron fake rate is shown in \textbf{(d)}, the neutral hadron fake rate in \textbf{(e)}, the photon fake rate in \textbf{(f)}.
  For efficiency and fake rate, we show the binomial statistical uncertainties from limited samples.
  We show the \ptmomentum response of charged hadrons in \textbf{(g)}, and the energy response of neutral hadrons in \textbf{(h)} and photons in \textbf{(i)}.

  }
  \label{fig:singleparticle_eff_fake_res}
\end{figure*}

First, we study the performance of the chosen \gls{gnn} model on single particle gun samples that were never used in training.
In \cref{fig:singleparticle_eff_fake_res}, we see that the performance of the baseline \gls{pf} and our proposed \gls{mlpf} algorithm is broadly similar for charged hadrons, while the efficiency and fake rate for neutral hadrons and photons are higher for \gls{pf}, especially at low calorimeter cluster energies.
Apart from differences in acceptance thresholds that we observe here, we do not expect or observe significant physical differences on single particle gun samples.
The \ptmomentum and energy response distributions are broadly similar for all particle types.
From this, we conclude that while not perfect, the baseline \gls{pf} algorithm is reasonably well tuned and represents a meaningful comparison point.

\begin{figure*}[!h]
  \centering
  \caption{The generated (truth) and reconstructed kinematic distributions for baseline particle flow (PF) and the proposed machine-learned particle flow (MLPF) algorithm.}
  \includegraphics[width=0.95\textwidth]{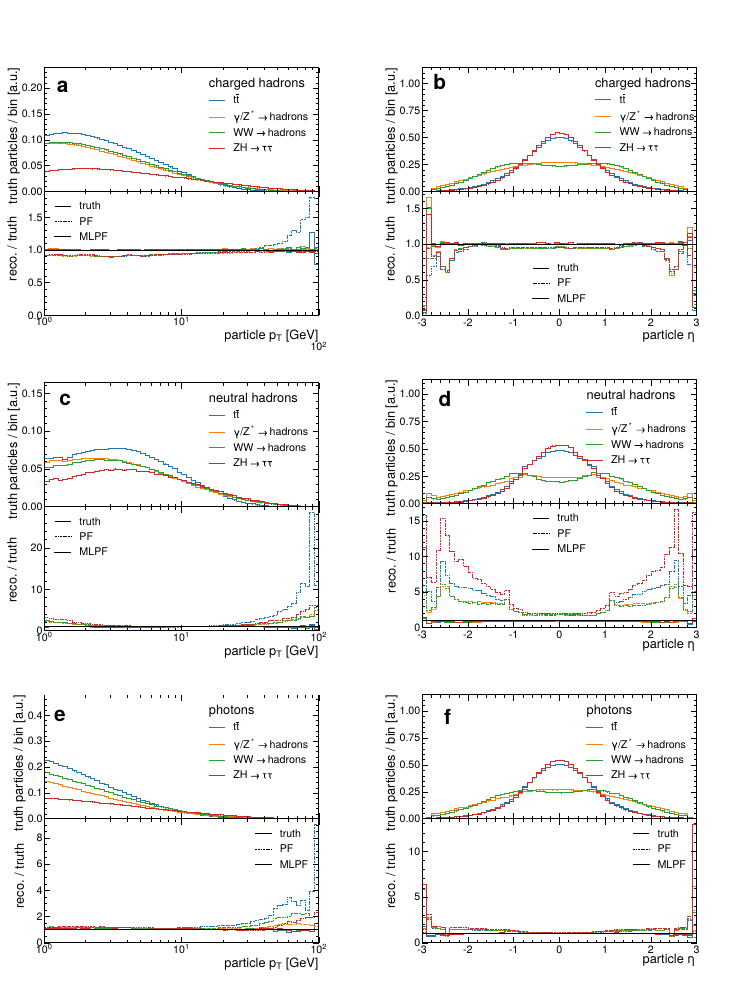}
  \label{fig:particles1}
  \caption*{The distribution of generated particles for \ttbar, \qq, \ww and \zh, and the ratio between the reconstructed and generated particle distributions for the baseline algorithm (dashed line) and the machine-learned algorithm (solid line) for charged hadrons in \textbf{(a)} and \textbf{(b)}, neutral hadrons in \textbf{(c)} and \textbf{(d)}, photons in \textbf{(e)} and \textbf{(f)}.}
\end{figure*}

\begin{figure*}[!h]
  \centering
  \caption{The generated (truth) and reconstructed kinematic distributions for baseline particle flow (PF) and the proposed machine-learned particle flow (MLPF) algorithm for electrons and muons.}
  \includegraphics[width=0.95\textwidth]{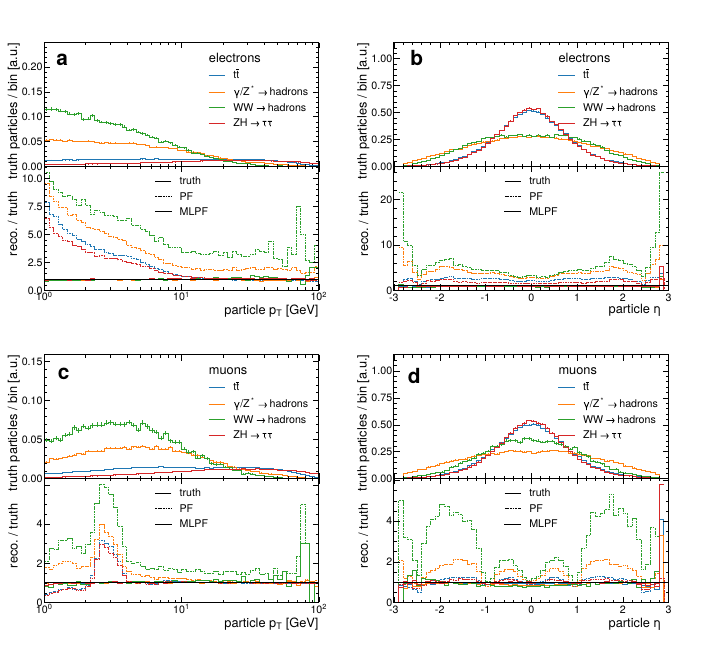}
  \caption*{The distribution of generated particles for \ttbar, \qq, \ww and \zh, and the ratio between the reconstructed and generated particle distributions for the baseline algorithm (dashed line) and the machine-learned algorithm (solid line) for electrons hadrons in \textbf{(a)} and \textbf{(b)} and muons in \textbf{(c)} and \textbf{(d)}.}
  \label{fig:particles2}
\end{figure*}

\begin{figure*}[!h]
  \centering
  \caption{The generated (truth) and reconstructed kinematic distributions for baseline particle flow (PF) and the proposed machine-learned particle flow (MLPF) algorithm for jets.}
  \includegraphics[width=0.95\textwidth]{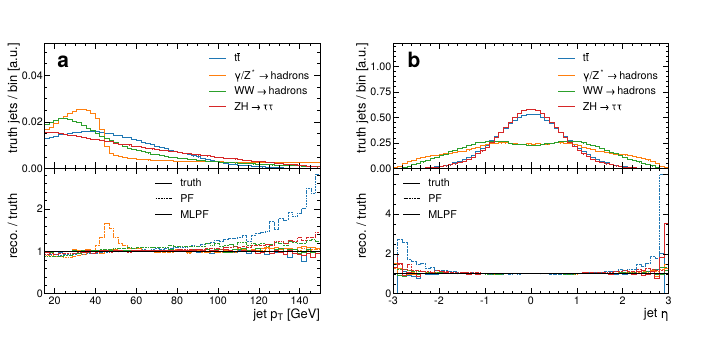}
  \caption*{The $\ptmomentum$ \textbf{(a)} and $\eta$ \textbf{(b)} distributions of generated jets for \ttbar, \qq, \ww and \zh, and the ratio between the reconstructed and generated jet distributions for the baseline algorithm (dashed line) and the machine-learned algorithm (solid line).}
  \label{fig:jets}
\end{figure*}

We also report the single-particle truth and reconstructed distributions for all held-out samples in \cref{fig:particles1,fig:particles2}, as well as for jets in \cref{fig:jets}.

\begin{figure*}[htpb]
    \centering
    \caption{Jet and total 3-momentum response in the validation samples, comparing the baseline particle flow (PF) and the machine-learned particle flow (MLPF).}
    \includegraphics[width=0.95\textwidth]{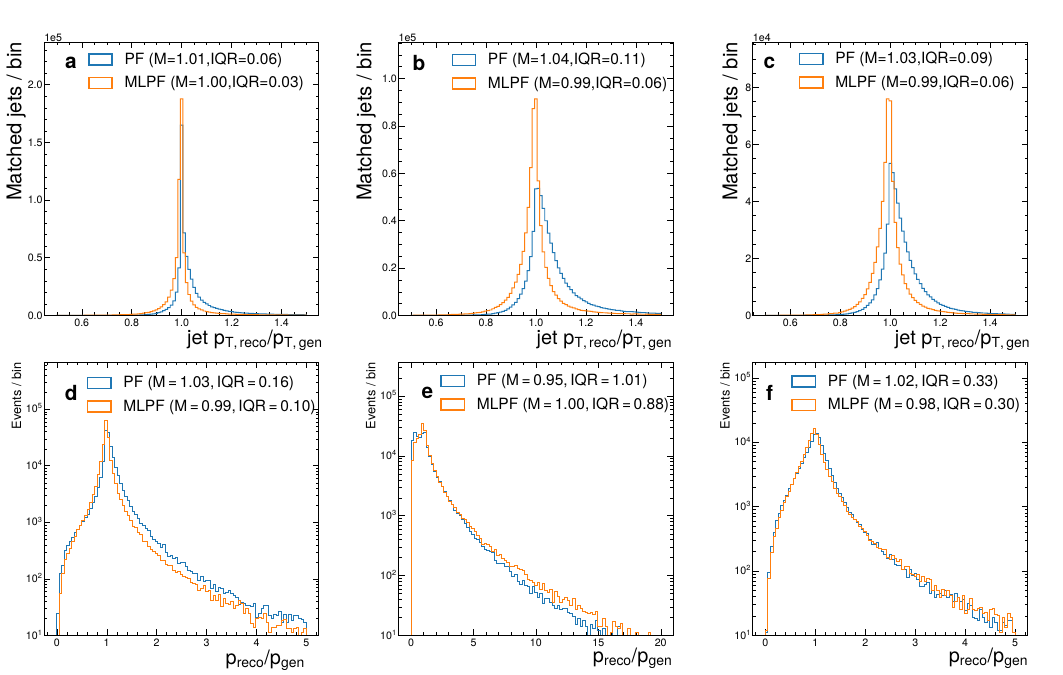}
    \caption*{The jet response for \zh samples in \textbf{(a)}, for \ww samples in \textbf{(b)} and for \ttbar with hadronic overlay on \textbf{(c)}. The total 3-momentum response is shown for \zh samples in \textbf{(d)}, for \ww samples in \textbf{(e)} and for \ttbar with hadronic overlay on \textbf{(f)}. We also show the median of the distribution ($M$), for which values closer to unity are better, and the interquartile range (IQR), for which lower values are better.}
    \label{fig:jet_mom_response_overall}
\end{figure*}

The physics performance in event-level quantities is summarized in~\cref{fig:jet_mom_response_overall}.
The jet response distribution of the baseline \gls{pf} algorithm is somewhat asymmetric, as the baseline jets are biased towards higher \ptmomentum values with respect to the matched generator jet, while the distribution from our proposed algorithm is significantly more symmetric around unity.
This is due to the overall momentum regression being optimized on the samples directly in \gls{mlpf}.

We use \zh, \ww and \ttbar PU10 events to evaluate out-of-distribution performance.
We evaluate the jet response by clustering reconstructed particles into jets, by matching the reconstructed and generator-level jets, and computing the ratio of reconstructed to generator-level jet \ptmomentum.
In all samples, the fraction of reconstructed jets from the \gls{gnn}-based \gls{mlpf} is the same or higher than for \gls{pf}, with generally an improved jet response width, quantified by the \gls{iqr} and median compared to the rule-based baseline.
This improvement over the baseline was not observed before hyperparameter tuning.
We also evaluate the total 3-momentum response for either \gls{pf} or \gls{mlpf} particles, and find that the \gls{mlpf} model improves both the median and \gls{iqr} of the total 3-momentum response distributions for all samples.

\begin{figure*}[htpb]
    \centering
    \caption{The jet and total 3-momentum resolution, comparing the baseline particle flow (PF) with the machine-learned algorithm (MLPF).}
    \includegraphics[width=0.95\textwidth]{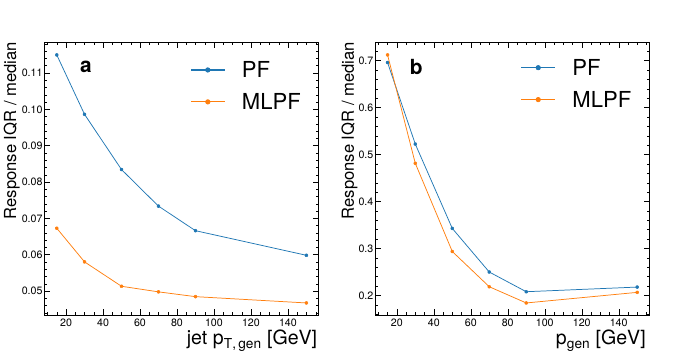}
    \caption*{The jet \ptmomentum response \textbf{(a)} and total 3-momentum response \textbf{(b)}, comparing the baseline particle flow (PF) with the machine-learned algorithm (MLPF), parameterised as the interquartile range (IQR) divided by the median  the \ttbar sample with hadronic overlay, evaluated in bins of generated jet \ptmomentum (total 3-momentum). Lower values correspond to better resolution.}
    \label{fig:jet_met_response_med_iqr}
\end{figure*}

To compare the momentum resolutions between the different approaches, while accounting for the differences in the momentum scales, we evaluate the metric of the \gls{iqr} divided by the median of the response distributions.
We quantify the evolution of this metric for the jet \ptmomentum (total event 3-momentum) in bins of generator-level jet \ptmomentum (generated total event 3-momentum) on the \ttbar with 10 PU interactions (\ttbar PU10) sample in~\cref{fig:jet_met_response_med_iqr}.
The baseline \gls{pf} and proposed \gls{mlpf} algorithms behave qualitatively similarly, with improved response \gls{iqr} over median values at higher generator-level jet \ptmomentum (total 3-momentum), while the \gls{mlpf} algorithm consistently outperforms the baseline \gls{pf} on this sample by up to 50\%.
This improvement has important implications for the sensitivity of key measurements at future colliders, such as those involving Higgs bosons that decay to bottom quark-antiquark pairs~\cite{Dawson:2022zbb}.

\begin{figure*}[htpb]
    \centering
    \caption{Computational performance of the baseline particle flow (PF) and the proposed machine-learned algorithm in inference and training.}
    \includegraphics[width=0.95\textwidth]{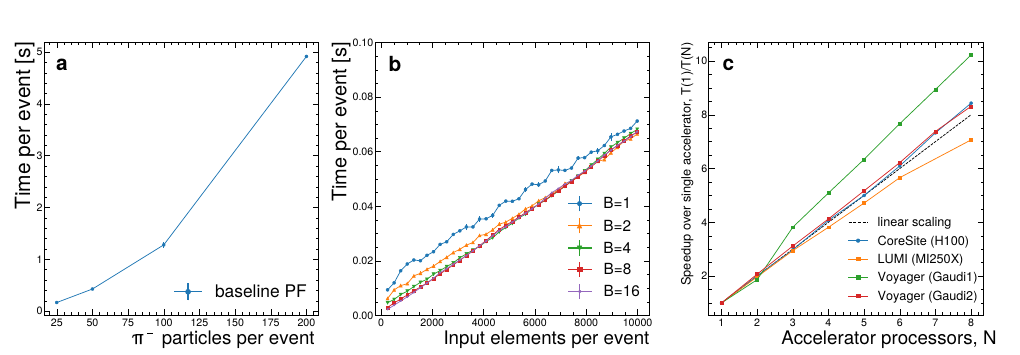}
    \caption*{Absolute timing of the baseline \gls{pf} \textbf{(a)} and the proposed \gls{gnn}-based algorithm \textbf{(b)}, illustrating the scaling with respect to input particle or element (track and cluster) multiplicity.
    The absolute processing time of the baseline algorithm on a single CPU thread is approximately 1,s/event at the reference point of 100 charged pions, which corresponds to approximately $96 \pm 3$ tracks and $170\pm20$ clusters. The runtime of the machine-learned particle flow (MLPF) algorithm on the Nvidia RTX2060S GPU at the reference point of $N=256$ elements, batch size $B=16$  is 2\,ms/event. This should not be interpreted as a complete, exhaustive and final computational benchmark, nor should it be used to claim 500-fold improvement in throughput from \gls{mlpf}, as the absolute timing of any algorithm is heavily dependent on optimizations and hardware. In \cite{Mokhtar:2023fzl}, the runtime of a heavily optimized version of the baseline algorithm was measured at 9\,ms/event, whereas the \gls{mlpf} model at 320\,ms/event, both on a single CPU thread.
    On panel \textbf{(c)}, we demonstrate the scaling of the training performance across multiple devices on a single machine on Nvidia, AMD, and Habana processor cards from the CoreSite, LUMI, and Voyager supercomputers, respectively.
    For the multi-device scaling test, Horovod~\cite{horovod} was used on CoreSite and Voyager, while the \TENSORFLOW \texttt{MirroredStrategy}~\cite{mirroredstrategy} was used on LUMI.
    The batch size was adjusted to fit a single device and the dataset was fully cached in RAM.}
    \label{fig:pf_timing}
\end{figure*}

An important factor in the development of scalable \gls{ml}-based full event reconstruction models is improving the computational throughput in future deployment scenarios.
Possible approaches to improve inference throughput could include more efficient model formulations and implementations~\cite{gu2023mamba}, sparsity and quantization.
Therefore, we compare the inference scalability of the rule-based \gls{pf} implementation on CPU and the proposed scalable \gls{gnn} implementation on GPU with respect to increasing input multiplicity, with the results summarized in~\cref{fig:pf_timing}.
Inference scaling tests of the \gls{gnn} model are tested on a small, 8\,GB consumer GPU (Nvidia RTX2060S) to emulate a resource-constrained scenario in edge deployment.

The rule-based version is run with 25, 50, 100 and 200 generated $\PGpm$ particles per event for 10 events.
We generated three different sets of these events varying the random seed each time, and measured the runtime of the particle-flow module in \texttt{Key4HEP}.
The runtime of the baseline increases nonlinearly with increasing particle multiplicity, and segmentation faults occur for more than approximately 200 particles per event, possibly due to the baseline code and configuration not being tuned for such a high number of particles.
Currently, it is only possible to track memory usage of the full baseline reconstruction chain, not individual algorithms.
However, we find that the maximum memory requirements increase approximately linearly from about 2\,GB for 25 particles to about 8\,GB for 200 particles.

The \gls{gnn} model is run with a varying input size on an 8\,GB consumer GPU multiple times to average over random fluctuations.
Both the batch size $B$, i.e., the number of batched events ($B \in \{1,2,4,8,16\}$) and the event size $N$, i.e., the number of input elements per event ($N = 256n$ for $n \in [1, 40]$) are varied independently.
The inference runtime scales approximately linearly with $N$.
We note that $B \gg 1$ is required to saturate the GPU, but this is highly model and device specific.
The maximum  GPU memory of our proposed \gls{mlpf} algorithm varies between about 300\,MB for the smallest tested configurations ($B=1, N=256$ to about 4.5\,GB for the largest tested configuration ($B=16, N=10240$), with allocations scaling in a stepwise manner due to the \gls{lsh} binning.

We confirm that the \gls{ml}-based reconstruction based on the \gls{lsh}-binned \gls{gnn} for full event reconstruction avoids the quadratic scaling present in a typical rule-based full event reconstruction algorithm.
It is likely that additional effort will be able to significantly improve the performance of both algorithms.
For example, in CMS, a $k$-dimensional tree~\cite{Bentley:kdtree} is used to avoid quadratic scaling~\cite{CMS:2017yfk}.
With aggressive quantization, the throughput of \gls{ml} models can be significantly improved with a negligible performance degradation~\cite{dettmers2022llm}.

At this point, it is not meaningful to compare the absolute throughput of the rule-based model on a CPU and the \gls{ml}-based model on a GPU, as neither method is particularly optimized for throughput, and the comparison is strongly affected by the specific choice of hardware.

The training scalability is also tested, with results presented in~\cref{fig:pf_timing}, on three different \gls{hpc} centers with different accelerator hardware: Nvidia H100 GPUs from Flatiron Institute's CoreSite cluster~\cite{coresite}, AMD MI250 GPUs from the LUMI supercomputer~\cite{lumi}, and Intel Habana Gaudi HPUs from the Voyager supercomputer~\cite{voyager}.
The LUMI supercomputer features GPU nodes with 64-core AMD Trento CPUs and four AMD MI250X cards, each card consisting of two accelerator chips.
Voyager is an NSF-funded supercomputer with 42 first-generation Intel Habana Gaudi (training) nodes, each with eight cards, two first-generation Intel Habana Goya (inference) nodes, a 400\,GbE Arista switch, and 3\,PB of Ceph file system available at the San Diego Supercomputer Center located at the University of California San Diego.
Intel Habana also provided additional access to eight Gaudi2 nodes in an HLS-Gaudi2 Deep Learning Server~\cite{gaudi2}.
For the training tests, there are some differences in the configuration on the HPCs.
For the AMD processors, multi-card training was implemented with a mirrored worker configuration, while for the Nvidia and Habana processors, Horovod~\cite{horovod} was used.
Events were zero-padded to a regular size of 512 elements per event.
A batch size of 250 events per device was used for the Nvidia, AMD, and Habana Gaudi2 processors, while a batch size of 100 per device was used for the Habana Gaudi processors.
We observe nearly linear scaling or better for all processors.
The scaling for the Habana Gaudi1 (Gaudi2) processors is enhanced by the all-to-all non-blocking intra-node network connection, where each processor has a 100 (300)\,Gb network connection to every other processor~\cite{voyager,gaudi2}.

\section{Conclusions}
\glslocalresetall
We have used a realistic simulation for the CLIC detector, conceptually similar to existing and future detector designs, to develop scalable machine learning models for full event particle flow reconstruction.
We compare two scalable \gls{ml} models: a \gls{gnn} model that uses \gls{lsh}, and a kernel-based transformer using large-scale \gls{hpo}.
Both of these models avoid quadratic scaling in memory and computation, thus allowing the processing of events from highly granular detectors.
We find that a \gls{gnn} model significantly outperforms the kernel-based transformer alternative and the baseline Pandora-based \gls{pf} on the basis of individual particles as well as event-level quantities such as jets and total 3-momentum.
Improved particle-level and event-level reconstruction can result in significant improvements for future flagship analyses such as those involving the Higgs boson decaying to bottom quarks.
At the same time, the flexible and learnable \gls{lsh} approach allows one to process complex events expected at future detectors with up to $10^4$ particles per event efficiently, while supporting various existing hardware accelerators such as Nvidia, AMD and Habana without additional porting effort.
Our work contributes to the existing body of research by proposing a new, challenging open dataset for particle flow reconstruction studies, by defining relevant benchmarks and by identifying efficient models that can solve these benchmarks better than existing rule-based algorithms without additional tuning.

Further research is possible in several directions.
First, it would be useful to repeat this exercise using the simulation from detectors that are taking data in Run 3 of the \gls{lhc}, e.g. in CMS, to study the performance of the reconstruction in more realistic conditions, and also study the reconstruction performance on real data.
Second, we are currently using a simple particle-based loss, while the use of contrastive-adversarial learning methods may allow one to account for event-level discrepancies more effectively~\cite{huang2023learning}.
Third, the hypergraph model~\cite{DiBello:2022iwf} shows promising physics performance but currently only supports small input sequences.
It may be interesting to extend the hypergraph construction over dynamically binned events using the \gls{lsh} approach.
While the current model works at the level of tracks and calorimeter clusters, our proposed approach scales naturally to cases where one considers the raw detector hits directly as an input, possibly allowing direct event reconstruction without having to tune clustering or tracking algorithms.
It may also be useful to construct features using semi-supervised or unsupervised learning from real data, to reduce the reliance on simulated datasets for supervised learning~\cite{Kishimoto:2023cys}.
Furthermore, large-context models are continuously improving, and it may be interesting to apply the latest developments such as \textsc{FlashAttention}~\cite{dao2022flashattention,dao2023flashattention2} on the models in this paper.
For improving throughput on e.g. CPUs, quantization has shown promise, and it may be interesting to investigate if this can be repeated for the type of models proposed for \gls{pf} reconstruction.
Finally, it is important to integrate the proposed \gls{ml}-based reconstruction models into reconstruction frameworks such as \textsc{CMSSW} and \textsc{Key4HEP}.

\section*{Data availability}
Our datasets are published following the findable, accessible, interoperable, and reusable principles.
The datasets are available in \cite{pata_joosep_2023_8260741}, the results including the trained model weight files in \cite{pata_joosep_2023_8328683}

\section*{Code availability}
Our code is published following the findable, accessible, interoperable, and reusable principles.
The code used for analysis is available in \cite{joosep_pata_2023_8290119}.

\section*{Authorship statement}
The authors contributed according to the contributor roles taxonomy (CRediT) categories as follows.
\textbf{Joosep Pata}: conceptualization, methodology, software, validation, data curation, writing (original draft), funding acquisition, resources.
\textbf{Eric Wulff}: methodology, software, data curation, writing (review and editing), resources.
\textbf{Farouk Mokhtar}: methodology, software, investigation, supervision.
\textbf{David Southwick}: software, investigation.
\textbf{Mengke Zhang}: validation, investigation.
\textbf{Maria Girone}: funding acquisition, resources.
\textbf{Javier Duarte}: conceptualization, software, resources, funding acquisition, supervision, writing (review and editing).

\section*{Acknowledgements}
Eric Wulff and David Southwick were supported by CoE RAISE.
The CoE RAISE project has received funding from the European Union's Horizon 2020 – Research and Innovation Framework Programme H2020-INFRAEDI-2019-1 under grant agreement no. 951733.

Joosep Pata was supported by the grant PSG864 of the Estonian Research Council.

Javier Duarte, Farouk Mokhtar, and Mengke Zhang were supported by U.S. Department of Energy (DOE) Award Nos. DE-SC0021187 and DE-SC0021396 (FAIR4HEP) and U.S. National Science Foundation (NSF) Cooperative Agreement OAC-2117997 (A3D3).
Farouk Mokhtar was also supported by a UCSD HDSI fellowship and an IRIS-HEP fellowship through NSF Cooperative Agreement OAC-1836650.
The use of Intel Habana processors and additional support for Javier Duarte was facilitated by NSF Award No. 2005369 (Voyager).
We also thank Amit Majumdar, Mahidhar Tatineni, Paul Rodriguez, and Marty Kandes at the San Diego Supercomputer Center and Leonid Goldgeisser, Charles Chen, Chen Levkovich, Leon Si Tran, and Chenna Bayapureddy at Intel Habana for technical support and assistance in producing results for the Gaudi2 processors.

The authors gratefully acknowledge the computing time granted through JARA on the supercomputer JURECA~\cite{jureca} at Forschungszentrum J\"{u}lich.
We acknowledge the Estonian Scientific Compute Infrastructure and NICPB for awarding this project access to the LUMI supercomputer, owned by the EuroHPC Joint Undertaking, hosted by CSC (Finland) and the LUMI consortium through Estonian Scientific Compute Infrastructure.

Datasets and software published following the findable, accessible, interoperable, and reusable (FAIR) principles support open, transparent, and reusable research, and therefore we thank our colleagues at the relevant collaborations for releasing the \texttt{Key4HEP}, \texttt{Marlin}, \texttt{EDM4HEP} and \texttt{CLICdp} software to the wider community, and encourage the wider adoption of the FAIR principles.

We are grateful to the reviewers and the editor for their careful reading of the manuscript and their comments which significantly improved the clarity of this paper.

We would also like to thank Nilotpal Kakati and Etienne Dreyer for cross-checks of our published code and datasets.

\section*{Competing interests}
The authors declare no competing interests.

\clearpage
\bibliography{references}

\end{document}